\documentclass[graybox]{svmult}

\usepackage{helvet}         
\usepackage{courier}        
\usepackage{type1cm}        
\usepackage{makeidx}         
\usepackage{graphicx}        
\usepackage{multicol}        
\usepackage[bottom]{footmisc}

\usepackage{hyperref}
\hypersetup{
    colorlinks=true,
    anchorcolor=red,
    urlcolor=magenta,
    citecolor=red,
    linkcolor=cyan}
    
\usepackage{cite}

\usepackage{amssymb}

\usepackage{amsmath}	

\usepackage{mathrsfs}  

\def\scr{\mathscr}

\def\Oint{O^{(\rm{int})}_{[2]}}
\def\Oincov{O^{(\rm{int})}}
\def\bbS{{\mathbb S}}
\def\bbT{{\mathbb T}}
\def\AdS{{\mathrm{AdS}}}

\begin{document}

\title*{Ramond States of the D1-D5 CFT away from the free orbifold point}
\author{Andre Alves Lima, Galen M.~Sotkov and Marian Stanishkov}
\institute{Andre Alves Lima \at Department of Physics, Federal University of Esp\'irito Santo, 29075-900, Vit\'oria, Brazil, \email{andrealves.fis@gmail.com}
\and 
Galen M.~Sotkov \at Department of Physics, Federal University of Esp\'irito Santo, 29075-900, Vit\'oria, Brazil, \email{gsotkov@gmail.com}
\and Marian Stanishkov \at Institute for Nuclear Research and Nuclear Energy, Bulgarian Academy of Sciences, 1784 Sofia, Bulgaria \email{marian@inrne.bas.bg}}
\maketitle

\abstract{The free orbifold point of the D1-D5 CFT must be deformed with a scalar marginal operator driving it to the region in moduli space where the holographic supergravity description of fuzzball microstates becomes available. We discuss the effects of the deformation operator on the twisted Ramond ground states of the CFT by computing four-point functions. One can thus extract the OPEs of the deformation operator with these Ramond fields to find the conformal dimensions of intermediate non-BPS states and the relevant structure constants. We also compute the anomalous dimensions at second order in perturbation theory, and find that individual single-cycle Ramond fields are renormalized, while the full multi-cycle ground states of the $S_N$ orbifold remain protected at leading order in the large-$N$ expansion.}

\bigskip

\noindent 
Contribution to the proceedings of the
\\
 \emph{XIV International Workshop} \textsc{Lie Theory and its Applications in Physics}
\\
 20-26 June 2021, Sofia, Bulgaria.

\section{Introduction}

In the decoupling limit, the Type IIB supergravity solution for the bound state of a large number $N_1$ of D1-branes wrapped around a circle $\bbS^1$, and a large number $N_5$ of D5-branes wrapped around $\bbT^4 \times \bbS^1$ develops a ``throat'' geometry; that is, it becomes $\AdS_3 \times \bbS^3 \times \bbT^4$, where $\AdS_3$ and $\bbS^3$ have the same (large) radius fixed by the branes' charges. The holographic CFT dual to this spacetime, called the D1-D5 CFT, is a two-dimensional superconformal field theory with ${\cal N} = (4,4)$ SUSY, and R-symmetry group ${\rm SO(4)} \cong {\rm SU(2)}_L \times {\rm SU(2)}_R$ associated with $\bbS^3$. The superconformal algebra also has a global ``${\rm SO(4)}$'' symmetry associated with $\bbT^4$. The large central charge, fixed by the $\AdS_3$ radius, is $c = 6 N$, where $N = N_1 N_5 \gg 1$. See e.g.~\cite{David:2002wn} for a review.

The D1-D5 system is a well-known laboratory for the study of microscopic properties of black holes in String Theory. The SUGRA solution (when the branes carry  internal momentum along their common $\bbS^1$ direction) has a horizon whose area matches the counting of CFT states, as famously discovered by Strominger and Vafa \cite{Strominger:1996sh}. But more than ``state counting'', a precise holographic dictionary exists, between states in the D1-D5 CFT and specific non-singular, horizonless `semiclassical'  SUGRA solutions \cite{Mathur:2005zp,Skenderis:2008qn} which have the same charges as --- and thus look like --- the extremal supersymmetric black hole away from the would-be horizon, outside the $\AdS_3 \times \bbS^3$ throat. Details of this dictionary, and the search for new and more general (less symmetric) horizonless and non-singular `microstate geometries' are part of what has become known as the fuzzball program, an ongoing endeavor, see e.g.~ \cite{Bena:2015bea,Giusto:2015dfa,Mathur:2005zp,Rawash:2021pik,Skenderis:2008qn}.

The D1-D5 CFT dual to gravity solutions is strongly-coupled, but conjectured to live in the moduli space of a free SCFT on the symmetric orbifold $(\bbT^4)^N / S_N$. (See \cite{David:2002wn} and references therein.) Holographic computations are usually done in this free orbifold point, where results are often exact, and then rely on the existence of non-renormalization theorems for some BPS-protected objects such as NS chiral operators, as well as on explicit matching with bulk computations, in special cases where they can also be performed in parallel with their CFT counterpart. Meanwhile, motivated by the increasing scrutiny of the relation between individual states and geometries, the effect of the deformation of the free orbifold towards the strongly-coupled CFT has been given attention over the years, see e.g.~\cite{Avery:2010er,Guo:2019pzk,Keller:2019yrr}.

Here we report the results published in  \cite{Lima:2020nnx,Lima:2020boh,Lima:2020urq,Lima:2021wrz,Lima:2020kek},  on the fate of twisted Ramond ground states $|{\scr R}_{[g]}\rangle$, with an arbitrary twist $g \in S_N$, when the free CFT is deformed by a marginal scalar modulus $\Oint$. In terms of the action,
\begin{equation}
S_{\rm{int}} = S_{\rm{free}}  + \lambda\int \! {\rm d}^2z \ O^{(\rm{int})}_{[2]} (z, \bar z)  
\end{equation}
with a coupling $\lambda$. Known microstate geometries are dual to coherent superpositions of Ramond ground states, or to specific excitations thereof \cite{Giusto:2015dfa,Rawash:2021pik,Skenderis:2008qn}, providing 
one important motivation --- besides the fact that they are naturally very elementary objects --- for their study in the deformed theory. Although the actual D1-D5 CFT is strongly coupled, we work with conformal perturbation theory to order $\lambda^2$, which is the lowest possible order where one can detect lifting of dimensions of $|{\scr R}_{[g]}\rangle$. The analysis hinges upon us being able to compute specific four-point functions involving the Ramond states and the deformation modulus.

\section{Ramond ground states and their four-point functions with the deformation operator }

Specifically, to assess the anomalous dimensions in second-order perturbation theory, we must compute 
\begin{equation} \label{4fnint}
\Big\langle
{\scr R}^\dagger_{[g]} (\infty,\bar \infty) 
\;
\Oint(1, \bar 1) 
\;
\Oint (u , \bar u) 
\;
{\scr R}_{[g]} (0,\bar 0)
\Big\rangle 
\end{equation}
that is a four-point function of twisted operators. The Hilbert space of the orbifold is divided into twisted sectors,  created by the insertion of `bare-twist fields' $\sigma_g (z)$, which introduces a branch cut at the point $z$ such that, when a given field crosses it, $g \in S_N$ permutes the copies $I = \{1, \dots, N\}$. Twisted operators are excitations of the bare twists.
For example, $\Oincov_{(2)}$ is an excitation of a transposition $\sigma_{(2)}$, and in the twisted sector of the cyclic permutation $(1,\cdots,n) \in S_N$ there are two R-charged holomorphic Ramond ground states (see the Appendix for definitions)
 \begin{align} \label{Ramondn}
R^{\pm}_{(n)}(z) 
=
\exp \left( \pm \frac{i}{2n} 
\sum_{I = 1}^n \big[ \phi_{1,I} - \phi_{2,I} \big] 
\right) \sigma_{(1,\cdots, n)}(z) 
\end{align} 
differing by R-charge $j_\pm = \pm \frac12$, and both having conformal weight $h^{\rm R}_n = \frac{6n}{24}$, appropriate for the Ramond sector of a CFT with central charge $6n$.  To obtain $S_N$-invariant operators from (\ref{Ramondn}), we sum over  all elements in the conjugacy class $[n]$,
\begin{equation}\label{normSninvR}
R^{s}_{[n]} = \frac{1}{\sqrt{N! | {\rm Cent}(n)|}} \sum_{h\in S_N} R^{s}_{h (n) h^{-1}} .
\end{equation}
The normalization factor, featuring the order of the centralizer of the twist permutation, $|{\rm Cent}(n)| = (N-n)!n$, counts multiplicities of terms in the sum, to ensure that $R^{s}_{[n]}$ has the same normalization as each of its components in the r.h.s. More generally, any $g \in S_N$ can be decomposed as a product $g = \prod_n (n)^{N_n}$, of (disjoint) cyclic permutations of length $n$, characterized by a partition $[N_n ] = \{ N_n \in {\mathbb N}  \, | \, \sum_{n} n N_n = N \}$, and the conjugacy class $[g]$ is the set of all permutations with the same partition $[N_n ]$. Generic $S_N$-invariant operators for multi-cycle permutations can be constructed analogously to (\ref{normSninvR}), to depend only on $[N_n ]$. This applies to the Ramond ground states ${\scr R}_{[g]}$ of the \emph{full} orbifold, which are \emph{not} the fields $R^s_{[n]}$, but products 
\begin{equation}\label{RamondComp}
{\scr R}_{[N^{(s)}_n]}
=
\Big[ \prod_{s,n} ( R^{s}_{[n]} )^{N^{(s)}_n} \Big]
\quad
\text{for a partition} \quad
\sum_{n,s} N^{(s)}_n = N .
\end{equation}
The partition takes into account an SU(2) ``spin index'' $s = \pm, \dot 1, \dot 2$ besides the twist. Any field with this structure has the correct conformal weight $h^{\rm R} = \frac{N}{4}$ for a CFT with central charge $c = 6N$, and
R-charge $j = \sum_{s,n} {N^{(s)}_n} \, j_{s}$.

Connected correlators of twisted fields are associated to branched coverings of the Riemann sphere, whose genera are determined by the twist permutations via the Riemann-Hurwitz formula.
It can be shown that (\ref{4fnint}) factorizes into a sum of connected functions where the Ramond fields have at most \emph{two} cycles \cite{Lima:2021wrz}, 
\begin{equation}\label{main1}
A^{s_1s_2}_{n_1n_2}
=
\Big\langle 
\big[ R^{s_1}_{[n_1]}R^{s_2}_{[n_2]} \big]^\dagger (\infty,\bar \infty) 
\;
\Oint(1, \bar 1) 
\;
\Oint (u , \bar u) 
\;
\big[ R^{s_1}_{[n_1]}R^{s_2}_{[n_2]} \big] 
(0,\bar 0)
\Big\rangle ,
\end{equation}
with associated genus-zero covering surface. The covering is given by the map
\begin{equation}\label{coverm}
z(t) = \left( \frac{t}{t_1}\right)^{n_1} 
\left( \frac{t-t_0}{t - t_\infty} \right)^{n_2} 
\left( \frac{t_1-t_\infty }{t_1-t_0} \right)^{n_2} 
\end{equation}
such that, respectively, the twists $n_1$ at $z = 0, \infty$ are lifted to $t = 0, \infty$ on the covering; the twists $n_2$ at $z=0,\infty$ lift to $t = t_0, t_\infty$; and the twists 2 at $z = 1, u$ to $t = t_1, x$. The monodromies impose that $t_0,t_1,t_\infty$ are functions of $x$, and $u(x) \equiv z(x)$ is
\begin{equation}	\label{uxm}
u(x) = 
	x^{n_1-n_2}  
	( x+ \tfrac{n_1}{n_2})^{n_1+n_2}  
	(x-1)^{-n_1+n_2} 
	(x - 1 + \tfrac{n_1}{n_2})^{n_2-n_1}  .
\end{equation}
Following \cite{Lunin:2000yv,Lunin:2001pw,Pakman:2009zz} we can use the covering surface as a tool for dealing with monodromies and compute (\ref{main1}). The final result is
\begin{align}
\label{main}
\begin{split}
&
A^{s_1s_2}_{n_1n_2} (u,\bar u) 
=
\frac{\varpi(n_1 n_2)}{N^2}
\sum_{\frak a = \frak1}^{2 \max(n_1,n_2)} | A^{s_1s_2}_{n_1n_2}
( x_{\frak a}(u)) |^2  
\\
&
A^{s_1s_2}_{n_1n_2} (x) =
\frac{1}{16n_1^2}
\Big[ C_{s_1s_2} + x(x - 1 + \tfrac{n_1}{n_2}) \Big] 
\\
&\qquad
\times
\frac{
x^{1-n_1+n_2}
(x-1)^{1+n_1+n_2}
( x+ \frac{n_1}{n_2} )^{ 1 - n_1 - n_2 }
(x -1 + \frac{n_1}{n_2} )^{ 1 + n_1 - n_2 }
}{
( x + \frac{n_1-n_2}{2n_2} )^4
}
\end{split} 
\end{align}
We have written the overall factor $\varpi / N^2$ in the large-$N$ limit, but apart from this the result holds for finite $N$ as well.  The constant $\varpi$ depends only on $n_1n_2$, and $C_{s_1s_2}$ only on $n_1,n_2$ and $s_1,s_2$. They can be found in \cite{Lima:2021wrz}.
 
Eq.(\ref{main}) is written in terms of the functions $x_{\frak a}(u)$ that are the inverses of $u(x)$, and cannot be found in general, but can be expanded in the coincidence limits $u \to 0$ and $u \to 1$, to give the associated conformal data. 
For example, for $u \to 0$, we obtain the OPE
\begin{align*}
&
\Oincov_{(2)}(u,\bar u)
\
 \big[ R^{s_1}_{(n_1)}  R^{s_2}_{(n_2)} \big] (0,\bar 0) \, | \varnothing \rangle
\\
&\qquad\qquad
=
\frac{ 
\langle 
Y^{s_1s_2 \dagger}_{(n_1+n_2)} 
\Oincov_{(2)} 
[ R^{s_1}_{(n_1)}  R^{s_2}_{(n_2)} ]
\rangle
}{
|u|^{2 + \frac{n_1 + n_2}{2} - \Delta^{s_1s_2}_Y }
}
Y^{s_1s_2}_{(n_1+n_2)} (0,\bar 0) \, |  \varnothing \rangle
 +\cdots
\end{align*}
The operator $Y^{s_1s_1}_{(n_1+n_2)}$ results from $\Oincov_{(2)}$ joining the two single-cycle Ramond fields. Its dimension $\Delta^{s_1s_2}_Y$,  as well as the structure constant in the denominator, can be found from the expansion of (\ref{main}). For instance, when $n_1 = n_2 = n$ we find 
\begin{align*}
\Delta^{+-}_{Y} 
=
\Delta^{\dot1\pm}_{Y } 
=
\Delta^{\dot1\dot2}_{Y } 
=
\Delta^{\dot1\dot1}_{Y } 
&=
\frac1n + n ,
\quad
\Delta^{++}_{Y } 
=  \frac{2}{n}  +  n .
\end{align*}
In the limit $u \to 1$, we find the (symbolic) OPE $\Oint \times \Oint =  [1] + [\sigma_{[3]}]$, expected from the composition of the two transpositions in the twists. The channel with the conformal family of the identity $[1]$, is in fact used to fix the constants in (\ref{main}).

The correlator (\ref{4fnint}) is a sum of functions (\ref{main}) for every required pairwise combination of single-cycle Ramond fields. This sum has ``symmetry factors'' depending on the multiplicity of equivalent strands in the product (\ref{RamondComp}), i.e.~on the form of the partition $[N^{(s)}_n]$, but these factors are not dynamical and can be disregarded in the computation of the anomalous dimension below. We note that there can be contributions from functions with \emph{single-cycle} Ramond fields and a genus-one covering surface, but we focus on the genus-zero functions here.

\section{Away from the free orbifold point}

Conformal perturbation theory gives the dimension 
$\Delta^{\rm R}_{\rm (ren)} = h^{\rm R}_{\rm (ren)} + \tilde h^{\rm R}_{\rm (ren)}$
of ${\scr R}_{[g]}$ at order $\lambda^2$, once we perform the integral
\begin{equation*}
\Delta^{\rm R}_{\rm (ren)} = \frac{N}{2}  - \frac{\pi}{2} \lambda^2 \int\! {\rm d}^2u \, \Big\langle
{\scr R}^\dagger_{[g]} (\infty,\bar \infty) 
\,
\Oint(1, \bar 1) 
\,
\Oint (u , \bar u) 
\,
{\scr R}_{[g]} (0,\bar 0)
\Big\rangle  .
\end{equation*}
The integrand is a sum of functions of the type (\ref{main}). With a change of variables we find, with unimportant constants ${\scr A}$ and ${\scr B}$,
\begin{align}
\begin{split}
\int\! {\rm d}^2u \, A^{s_1s_2}_{n_1n_2} (u,\bar u)
&= \int\! {\rm d}^2x \, \big| u'(x) \,  A^{s_1s_2}_{n_1n_2} (x) \big|^2 
\\
&= 
{\scr A}
\int \! {\rm d}^2y\, | 1 - y |^{-3} 
+
{\scr B}
\int \! {\rm d}^2y\,  | 1 - y |^{-3} |y - w|^{2}
\end{split}
\end{align} 
which is divergent, but can be regularized by analytic continuation:
\begin{align*} 
 \int \! \frac{{\rm d}^2y}{ | y-1 |^{3} }
= \lim_{a \to 0} \frac{4 \pi }{ \Gamma(-a) } 
= 0 , 
\quad 
\int \! {\rm d}^2y \,  \frac{ |y - w|^{2} }{| 1 - y |^{3}} \sim \lim_{a \to 1} \frac{-16\pi}{ \Gamma(-a)}
= 0 
\end{align*} 
Hence
\begin{equation}
\Delta^{\rm R}_{\rm (ren)} = \tfrac{1}{2} N + \rm{O}(\lambda^3)
\end{equation}
at least in the large-$N$ limit, where only the genus-zero connected functions contribute to (\ref{4fnint}). The dimension of the Ramond ground states is protected.

\begin{figure}[t]
\sidecaption[t]
\begin{centering}
\includegraphics[scale=0.35]{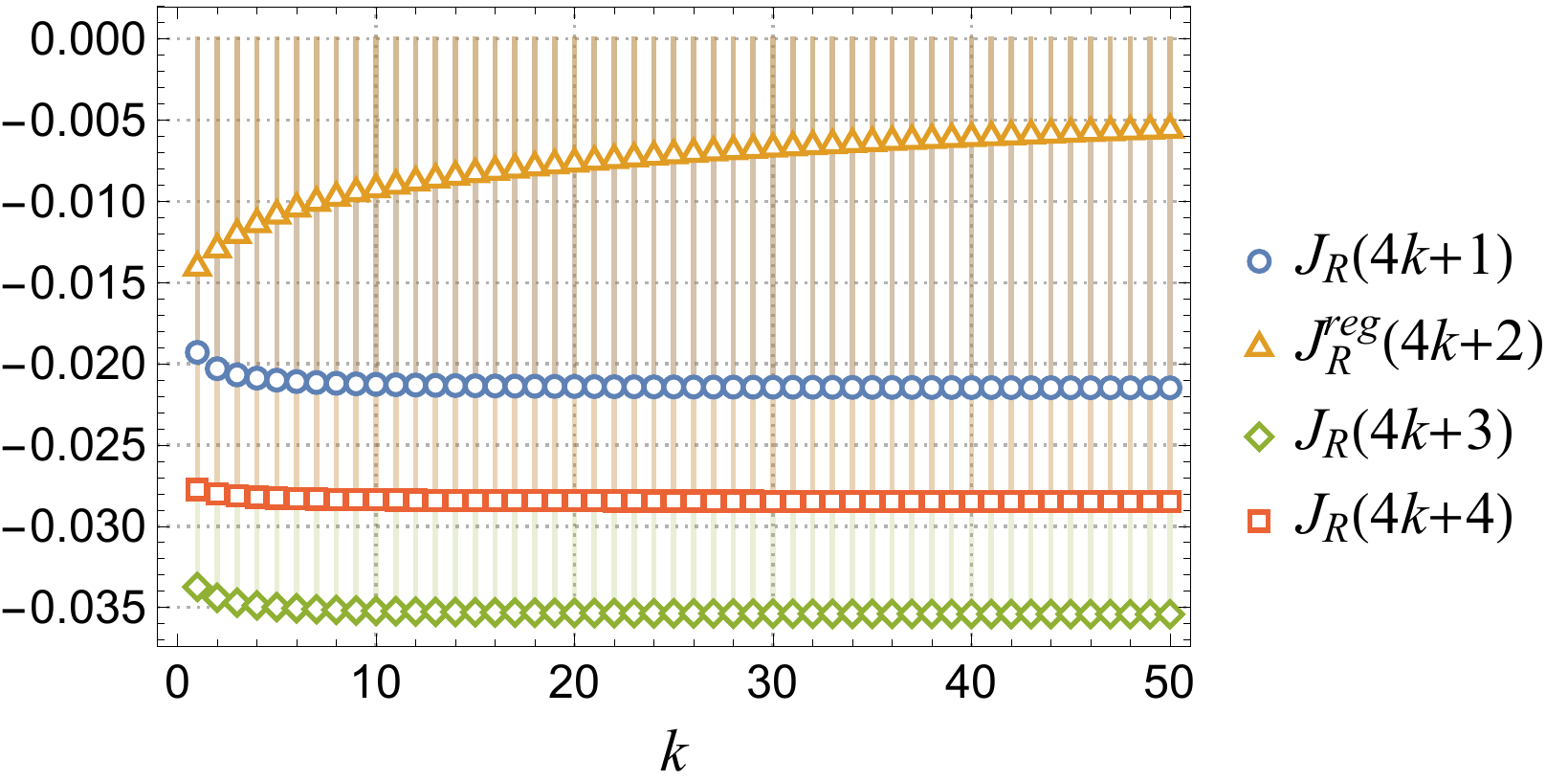}
\includegraphics[scale=0.48]{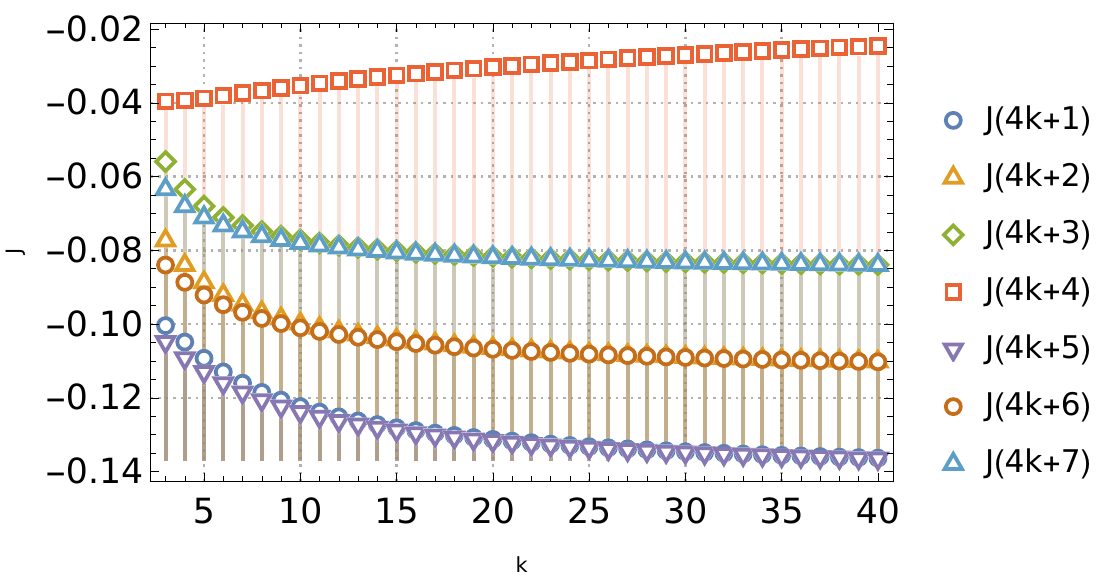}
\end{centering}
\caption{Plots of the integral $J_R(n)$ in Eq.(\ref{JR}), for  R-charged (left) and R-neutral (right) single-strand Ramond ground states of twist $n = 4k +\ell$.}
\label{Plots}      
\end{figure}

A curious feature of the non-renormalization is that it only occurs for the total orbifold ground states with $h^{\rm R} = \frac{1}{4} N$. Individual single-cycle states with $h^{\rm R}_n = \frac{1}{4} n$ \emph{are} renormalized.
We can see that by using similar methods to compute
\begin{equation}\label{JR}
J_R =
\int\! {\rm d}^2 u \
\Big\langle
R^{s\dagger}_{[n]}
(\infty, \bar \infty) 
\Oint(1,\bar 1) 
\Oint(u,\bar u) 
R^{s}_{[n]} (0,\bar 0)
\Big\rangle 
\end{equation}
whose numerical values are plotted in Fig.\ref{Plots}. The functions in the integrand contain a leading genus-zero part coming from $\Oint$ joining the field  $R^{s}_{(n)}$ with an untwisted vacuum strand, that is absent in the decomposition of the correlator (\ref{4fnint}).

In summary, we find that the $c = 6N$ CFT Ramond ground states
are protected, as it should be expected, since they are associated with SUGRA fuzzball solutions and related to BPS NS-chiral fields by spectral flow of the $c = 6N$ theory. However, the individual single-cycle components
$R^{s}_{[n]}$ do lift, and acquire anomalous dimensions when the CFT is driven away from the free orbifold point.

\section*{Appendix}

The $({\mathbb T}^4)^N/S_N$ orbifold
has $N$ copies of a `seed' ${\cal N} = (4,4)$ SCFT with central charge $c_{\rm seed} = 6$. The total central charge is $c = 6 N$. Each copy, labeled by an index $I = \{1,\dots,N\}$, has 4 real bosons and $(4 + \tilde4)$ real fermions, all free, which can be gathered into SU(2) doublets. The holomorphic Ramond fields in the text are constructed from the bosonized fermions 
$\psi^{\alpha\dot 1}_I (z) = [ e^{- i \phi_{2,I}(z)}  ,  e^{- i \phi_{1,I}(z)} ]^T$
and
$\psi^{\alpha\dot 2}_I (z) = [ e^{ i \phi_{2,I}(z)}  ,  -  e^{i \phi_{1,I}(z)} ]^T$.
The indices $\alpha= \pm$ correspond to the holomorphic R-symmetry group $\rm{SU}(2)_L$ and $\dot A = \dot 1, \dot 2$ to the factor ${\rm SU}(2)_2$ of the global symmetry.

\bigskip

\noindent
{\bf Acknowledgements.}
This work was partially supported by the Bulgarian NSF grants KP-06-H28/5 and KP-06-H38/11.

\end{document}